\documentclass[aip,jmp,unsortedaddress,12pt]{revtex4-1}

\usepackage{amsthm,amsfonts,amssymb,amsmath,dsfont}
\usepackage[T1]{fontenc}
\usepackage{graphicx,graphics,color}
\usepackage{colortbl}

\usepackage{natbib}

\usepackage[pdftex,colorlinks=true,linkcolor=blue]{hyperref}

\begin{document}

\newcommand{\nn}{\nonumber}
\newcommand{\Tr}{\mathrm{Tr}}
\newcommand{\Ln}{\mathrm{Ln}}
\newcommand{\bra}{\langle}
\newcommand{\ket}{\rangle}
\newcommand{\del}{\partial}
\newcommand{\vt}{\vec}
\newcommand{\dg}{^{\dag}}
\newcommand{\cg}{^{*}}
\newcommand{\vep}{\varepsilon}
\newcommand{\suml}{\sum\limits}
\newcommand{\prodl}{\prod\limits}
\newcommand{\intl}{\int\limits}
\newcommand{\til}{\widetilde}
\newcommand{\mcl}{\mathcal}
\newcommand{\mfk}{\mathfrak}
\newcommand{\ds}{\displaystyle}
\renewcommand{\dim}{\mathrm{dim}}
\renewcommand{\Re}{\mathrm{Re}}
\renewcommand{\Im}{\mathrm{Im}}
\renewcommand{\mod}{\mathrm{\,mod\,}}
\renewcommand{\b}{\overline}

\title{Semiclassical propagator for $\mathrm{SU}(n)$ coherent states}

\author{Thiago F. Viscondi}
\email{viscondi@ifi.unicamp.br}
\author{Marcus A. M. de Aguiar}
\affiliation{Instituto de Física ``Gleb Wataghin'', Universidade Estadual de Campinas, 13083-859, Campinas, SP, Brazil}

\date{\today}

\begin{abstract}
We present a detailed derivation of the semiclassical propagator in the $\mathrm{SU}(n)$ coherent state representation. In order
to provide support for immediate physical applications, we restrict this work to the fully symmetric irreducible representations,
which are suitable for the treatment of bosonic dynamics in $n$ modes, considering systems with conservation of total particle
number. The derivation described here can be easily extended to other classes of coherent states, thus representing an
alternative approach to previously published methods.
\end{abstract}



\maketitle

\section{Introduction}
\label{sc:intro}

The coherent states are a fundamental tool in quantum mechanics. They were initially envisioned by
Schrödinger\cite{Schrodinger26} as minimum uncertainty Gaussian states whose dynamics has maximum similarity to the classical
oscillator. The interest in these states, which are related to the Weyl-Heisenberg algebra, was again aroused by the work of
Glauber\cite{Glauber63a, Glauber63b, Glauber63c}, Klauder\cite{Klauder60, Klauder63a, Klauder63b} and
Sudarshan\cite{Sudarshan63}, with the emergence of the first applications directed to quantum optics and semiclassical
approximations.

The generalization of the coherent states to arbitrary Lie groups was initially studied by Klauder, but a complete and detailed
definition of the generalized coherent states was only later developed by Perelomov\cite{Perelomov72} and
Gilmore\cite{Gilmore72}. In this way, important properties of the harmonic-oscillator coherent states were extended to other
algebras of  physical interest, resulting in an ideal setting for the study of the quantum-classical
correspondence\cite{Zhang90}.

Although the semiclassical methods also have their origin in the early days of quantum mechanics, the first calculations
concerning semiclassical propagators in the representation of harmonic-oscillator coherent states were only developed in the last
decades of the past century\cite{Weissman82, Weissman83, Huber87, Huber88}, as an attempt to transpose the practical difficulties
of semiclassical propagators in the coordinate and momentum representations.

However, similarly to the very definitions of coherent states, the generalization of semiclassical propagators for arbitrary
groups occurred only in the decades following the semiclassical approximations with Gaussian states. In particular, aiming at the
treatment of systems with spin, various correct derivations of the $\mathrm{SU}(2)$ semiclassical propagator\cite{Solari87,
Vieira95, Kochetov95a} were obtained independently. Amidst these different methods, Kochetov\cite{Kochetov95b,Kochetov98}
generalized his results to arbitrary Lie groups.

The study of semiclassical propagators is motivated mainly by the computational gain in relation to exact quantum calculations,
considering practical applications in systems with many degrees of freedom or large quantum numbers, in which cases it is
believed that the semiclassical approximation is efficient and accurate. From the less pragmatic point of view, the use of
classical quantities for the calculation of quantum corrections allows us a greater understanding of the quantum-classical
correspondence principle, since it establishes more precisely the boundaries and intersections between the two theories.

We propose in this paper an independent approach for the derivation of the semiclassical propagator in arbitrary coherent state
representations. However, to make this text clearer and more accessible, here we exemplify our method only for the coherent
states related to the fully symmetric representations of the $\mathrm{SU}(n)$ group\cite{Gilmore75}. These coherent states are
appropriate for the treatment of bosonic systems with a finite number of modes and conservation of the total particle number.
Therefore, the $\mathrm{SU}(n)$ coherent states are important for several physical applications, as for example, the research on
Bose-Einstein condensates in multi-well traps\cite{Viscondi09, Trimborn08, Trimborn09}.

The major technical difficulty encountered in the derivation of semiclassical propagators is related to the explicit calculation
of path integrals for quadratic approximations of the action functional. The topology and dimensionality of the classical phase
space can make this task quite complex. Recently, Braun and Garg\cite{Braun07a, Braun07b} developed a simple and transparent
solution to the semiclassical propagator for an arbitrary number of degrees of freedom, but considering only the direct product
of many harmonic-oscillator or $\mathrm{SU}(2)$ coherent states. Using a slightly modified form of their previous findings and a
suitable transformation of variables, we extend the path integral calculation to the $\mathrm{SU}(n)$ coherent states.

The remainder of this paper is divided as follows: in large part of section \ref{sc:harmonic} we reproduce the main results of
Braun and Garg\cite{Braun07a} concerning the semiclassical propagator for multidimensional harmonic-oscillator coherent states.
At the end of this section, we reformulate the identity that relates the path integral resulting from the semiclassical
approximation to dynamical quantities calculated on the classical trajectory. This fundamental result is used in section
\ref{sc:pathint}, where we make a detailed derivation of the semiclassical propagator in the $\mathrm{SU}(n)$ coherent state
representation. Finally, the section \ref{sc:conclu} is intended to our concluding remarks.

\section{Semiclassical propagator for $(n-1)$ harmonic modes}
\label{sc:harmonic}

\subsection{Harmonic-oscillator coherent states}
\label{ssc:cohstat}

The coherent state for $(n-1)$ harmonic modes is given by:

\begin{equation}
  |z\ket=e^{-\frac{z\cg
  z}{2}}\suml_{m_{1},m_{2},\ldots,m_{n-1}=0}^{\infty}
  \left(\prodl_{j=1}^{n-1}\frac{z_{j}^{m_{j}}}{\sqrt{m_{j}!}}\right)|m_{1},m_{2},\ldots,m_{n-1}\ket;
  \label{eq1p1}
\end{equation}

\noindent where $\{|m_{1},m_{2},\ldots,m_{n-1}\ket\}$ is the usual basis of the bosonic Fock space $\mathds{B}^{n-1}$ for
$\displaystyle{(n-1)}$ modes, such that $m_{j}$ is the occupation in the $j$-th mode. The complex vector $z =\left(z_{1}, z_{2},
\ldots, z_{n-1}\right)^{T}$, with $(n-1)$ entries, parameterizes the entire set of coherent states. Also, according to the
adopted notation, the juxtaposition of two vectors $a$ and $b$ represents the matrix product
$ab=a_{1}b_{1}+a_{2}b_{2}+\ldots+a_{n -1}b_{n-1}$.

The coherent states \eqref{eq1p1} constitute an overcomplete set in Hilbert space, which enables us to write the following diagonal
resolution of the identity in $\mathds{B}^{n-1}$:

\begin{equation}
  \intl_{z\in \mathds{C}^{n-1}} d\mu(z\cg,z)\;|z\ket\bra z|=\mathds{1};\qquad
  d\mu(z\cg,z)=\prodl_{j=1}^{n-1}\frac{d^{2}z_{j}}{\pi};
  \label{eq1p2}
\end{equation}

\noindent where $d^{2}z_{j}=dq_{j}dp_{j}=\frac{dz_{j}dz\cg_{j}}{2i}$, with $q_{j} = \Re\left(z_{j}\right)$ and $p_{j} =
\Im\left(z_{j}\right)$.

\subsection{Coherent state propagator}
\label{ssc:cohstatpropag}

The quantum propagator in the coherent state representation is defined as the matrix element of the time evolution operator
between the initial coherent state $|z_{i}\ket$ and the final coherent state $|z_{f}\ket$, assuming a propagation period $\tau$:

\begin{equation}
  K(z_{f}\cg,z_{i};\tau)=\bra z_{f}|e^{-iH\tau}|z_{i}\ket.
  \label{eq1p3}
\end{equation}

In the above definition, for simplicity, we assumed $\hbar=1$ and the temporal independence of the Hamiltonian $H$. However, the
main results presented in this work, including the semiclassical approximation of the propagator, are also valid for Hamiltonians
with explicit time dependence\cite{Baranger01}. Expanding the propagator \eqref{eq1p3} to second order around the classical
trajectory, which is defined later, we obtain the following form for its semiclassical approximation:

\begin{equation}
  K_{sc}(z\cg_{f},z_{i};\tau)=
  e^{iS_{c}-\frac{1}{2}(|z_{i}|^{2}+|z\cg_{f}|^{2})}K_{red}(z\cg_{f},z_{i};\tau);
  \label{eq1p4}
\end{equation}

\noindent where $S_{c}$ is the action functional $S$ calculated on the classical trajectory, and $K_{red}$ is the reduced propagator,
which introduces quantum corrections. The classical trajectories, by definition, are the extremes of the action functional:

\begin{equation}
  iS(z\cg_{f},z_{i};\tau)=
  \intl_{0}^{\tau}\left\{\frac{1}{2}(\dot{\b{z}}z-\b{z}\dot{z})-i\mcl{H}(\b{z},z)\right\}dt
  +\frac{1}{2}[z\cg_{f}z(\tau)+\b{z}(0)z_{i}].
  \label{eq1p5}
\end{equation}

In the previous equation, to emphasize the independence of the variables $z$ and $z\cg$, we made the change of notation
$z\cg\rightarrow\b{z}$. Also in \eqref{eq1p5}, we defined the function $\mcl{H}(\b{z},z)=\frac{\bra \b{z}\cg|H|z\ket}{\bra
\b{z}\cg|z\ket}$, which represents the effective classical Hamiltonian. The extremization of the action functional implies the
following classical equations of motion\cite{Note1}:

\begin{equation}
  \dot{z}=-i\frac{\del\mcl{H}}{\del\b{z}};\qquad
  \dot{\b{z}}=i\frac{\del\mcl{H}}{\del z}.
  \label{eq1p6}
\end{equation}

According to \eqref{eq1p3}, the correct classical trajectory must respect the boundary conditions $z(0)=z_{i}$ and
$\b{z}(\tau)=z\cg_{f}$. Generally, if we consider the variables $\b{z}$ as the complex conjugates of $z$, then the boundary
conditions would make the system of differential equations \eqref{eq1p6} overdetermined, since the two vector equations would
become redundant. Therefore, for arbitrary classical trajectories, $\b{z}(t)$ is not the complex conjugate of $z(t)$, henceforth
denoted by $z\cg(t)$, except for a specific boundary condition in which accidentally $\b{z}(0)=z_{i}\cg$.

The reduced propagator corresponds to an infinite dimensional integral of the second variation of the action $\delta^{2}S_{c}$:

\begin{equation}
  K_{red}(z\cg_{f},z_{i};\tau)=
  \int D\mu(\b{\eta},\eta)\exp\left\{\frac{i}{2}\delta^{2}S_{c}\right\};\quad
  D\mu(\b{\eta},\eta)=\lim_{M\rightarrow\infty}\prodl_{k=1}^{M-1}
  \left[\prodl_{j=1}^{n-1}\frac{d^{2}\eta_{j}^{k}}{\pi}\right].
  \label{eq1p7}
\end{equation}

The integration variables $\eta^{k}=z^{k}-z^{k}_{c}$ and $\b{\eta}^{k}=\b{z}^{k}-\b{z}^{k}_{c}$ represent the departure from the
classical trajectory, denoted by $z_{c}^{k}$ and $\b{z}_{c}^{k}$, at the time $t_{k}=k\vep$, for $k=0,1,2,\ldots,M$ and $\vep =
\frac{\tau}{M}$. These variables are subject to the boundary conditions $\eta^{0}=\eta(0)=0$ and $\b{\eta}^{M} = \b{\eta}(\tau) =
0$, also in agreement with the propagator \eqref{eq1p3}.

By linearization of the equations \eqref{eq1p6}, we can also obtain the classical equations of motion for deviations from the
classical trajectory. In this way we define the matrices $R_{jk}$, which are essential for later results\cite{Note2}:

\begin{equation}
  \left(\begin{array}{c} \delta\dot{z}(t) \\ \delta\dot{\b{z}}(t) \end{array}\right)
  =\left(\begin{array}{c c} -i\frac{\del^{2}\mcl{H}}{\del\b{z}\del{z}} &
  -i\frac{\del^{2}\mcl{H}}{\del\b{z}^{2}} \\
  i\frac{\del^{2}\mcl{H}}{\del{z}^{2}} &
  i\frac{\del^{2}\mcl{H}}{\del{z}\del\b{z}} \end{array}\right)
  \left(\begin{array}{c} \delta z(t) \\ \delta\b{z}(t) \end{array}\right)
  =\left(\begin{array}{c c} R_{11} & R_{12} \\ R_{21} & R_{22} \end{array}\right)
  \left(\begin{array}{c} \delta z(t) \\ \delta\b{z}(t) \end{array}\right).
  \label{eq1p9}
\end{equation}

The quantities in the previous equation are calculated on the classical trajectory. Notice that $\delta z$ and $\delta\b{z}$,
which represent the dynamical variables of equation \eqref{eq1p9}, deserve different notation from the integration variables
$\eta^{k}$ and $\b{\eta}^{k}$ due to their fundamentally unrelated purposes and definitions.

After a lengthy derivation, the result of the Gaussian integral \eqref{eq1p7} can be written in terms of dynamical quantities
calculated on the classical trajectory\cite{Note3}:

\begin{equation}
  K_{red}(z\cg_{f},z_{i};\tau)
  =e^{\frac{1}{2}\intl_{0}^{\tau}\Tr\left[R_{22}(t)\right]dt}
  \sqrt{\det\left[\frac{\del\b{z}(0)}{\del\b{z}(\tau)}\right]}.
  \label{eq1p10}
\end{equation}

The matrix $\frac{\del\b{z}(0)}{\del\b{z}(\tau)}$ represents the response of the ``free'' trajectory extreme $\b{z}(0)$ to a
change in the boundary condition $\b{z}(\tau)=z\cg_{f}$. Employing the functional \eqref{eq1p5}, we can show the identity
$\frac{\del\b{z}(0)}{\del\b{z}(\tau)}=i\frac{\del^{2}S_{c}}{\del z_{i}\del z\cg_{f}}$, relating the second factor of
\eqref{eq1p10} to a second derivative of the action.

\subsection{Continuous limit of $\delta^{2}S_{c}$}
\label{ssc:contlimit}

In the original text of Braun and Garg\cite{Braun07a}, only the time discretized form of $\delta^{2}S_{c}$ was employed in order
to obtain the result \eqref{eq1p10}. In this approach, the Gaussian integrals in \eqref{eq1p7} are calculated sequentially,
before the application of the continuous-time limit ($\vep\rightarrow0$, $M\rightarrow\infty$ and $\vep M=\tau$). However, we can
write the continuous form of $\delta^{2}S_{c}$ directly from \eqref{eq1p5}:

\begin{equation}
  i\delta^{2}S_{c}=\intl_{0}^{\tau}dt\left\{(\dot{\b{\eta}}\eta-\b{\eta}\dot{\eta})
  -\eta R_{21}\eta-\eta R_{22}\b{\eta}+\b{\eta}R_{11}\eta+\b{\eta}R_{12}\b{\eta}\right\}.
  \label{eq1p11}
\end{equation}

Substituting \eqref{eq1p11} in \eqref{eq1p7}, we obtain a new expression for the reduced propagator:

\begin{equation}
  K_{red}(z\cg_{f},z_{i};\tau)=\int D\mu(\b{\eta},\eta)\;\exp\left\{
  \frac{1}{2}\intl_{0}^{\tau}dt\left[(\dot{\b{\eta}}\eta-\b{\eta}\dot{\eta})
  -\eta R_{21}\eta-2\eta R_{22}\b{\eta}+\b{\eta}R_{12}\b{\eta}\right]
  \right\};
  \label{eq1p12}
\end{equation}

\noindent where we used the identity $R_{22}^{T}=-R_{11}$. Comparing the representations \eqref{eq1p10} and \eqref{eq1p12} of
$K_{red}$, we can write a formal identity between the path integral of $\delta^{2}S_{c}$ in its continuous form and the
corresponding solution described by classical quantities:

\begin{equation}
  \int D\mu(\b{\eta},\eta)\;e^{
  \frac{1}{2}\intl_{0}^{\tau}dt\left[(\dot{\b{\eta}}\eta-\b{\eta}\dot{\eta})
  -\eta R_{21}\eta-2\eta R_{22}\b{\eta}+\b{\eta}R_{12}\b{\eta}\right]}
  =e^{\frac{1}{2}\intl_{0}^{\tau}\Tr\left[R_{22}(t)\right]dt}
  \sqrt{\det\left[\frac{\del\b{z}(0)}{\del\b{z}(\tau)}\right]}.
  \label{eq1p13}
\end{equation}

We must always remember that the previous expression is associated with the boundary conditions $\eta(0)=0$ and $\b{\eta}(\tau)=0$.
The equation \eqref{eq1p13} is the main result of this section, since its generalization to $\mathrm{SU}(n)$ coherent states is
almost straightforward, as shown below.

\section{Path integral for $\mathrm{SU}(n)$ coherent states}
\label{sc:pathint}

\subsection{$\mathrm{SU}(n)$ coherent states}
\label{ssc:suncohstat}

The coherent states related to the \textit{fully symmetric irreducible representations} of $\mathrm{SU}(n)$ are given
by\cite{Gilmore75}:

\begin{equation}
  |N;w\ket=\suml_{m_{1}+m_{2}+\ldots+m_{n}=N}
  \left(\frac{N!}{m_{1}!m_{2}!\ldots m_{n}!}\right)^{\frac{1}{2}}
  \left(\prodl_{j=1}^{n-1}w_{j}^{m_{j}}\right)
  \frac{|m_{1},m_{2},\ldots,m_{n}\ket}{(1+w\cg w)^{\frac{N}{2}}}.
  \label{eq2p1}
\end{equation}

Here again $\{|m_ {1},m_{2},\ldots,m_{n}\ket\}$ is the usual basis of the bosonic Fock space $\mathds{B}_{N}^{n}$ for $n$ modes
and $N$ particles, such that $m_{j}$ is the occupation of the $j$-th mode. Similarly to the vector $z$ defined in equation
\eqref{eq1p1}, the number of entries of the complex vector $w=\left(w_ {1},w_ {2},\ldots,w_{n-1}\right)^{T}$ is $ (n-1) $.
However, the number of bosonic modes is now $n$, but with occupations restricted by $\suml_{j=1}^{n}m_{j}=N$. Note that the total
number of particles $N$ is the only index needed to uniquely specify a fully symmetric irreducible representation of
$\mathrm{SU}(n)$.

The $\mathrm{SU}(n)$ coherent states are suitable for the treatment of bosonic systems with a finite number of modes and
conservation of the total particle number. Therefore, henceforth we consider a system with a fixed number of bosons, so that the
evolution of its state is restricted to only one $\mathrm{SU}(n)$ irreducible representation. Thus, we also simplify the notation
by making $|N;w\ket\rightarrow|w\ket$.

Just as in the harmonic-oscillator case, the coherent states in \eqref{eq2p1} are normalized, but they are not orthogonal. The overlap
between two $\mathrm{SU}(n)$ coherent states is given by:

\begin{equation}
  \bra w'|w\ket=\frac{(1+w'^{*} w)^{N}}{(1+w'^{*} w')^{\frac{N}{2}}(1+w\cg w)^{\frac{N}{2}}}.
  \label{eq2p2}
\end{equation}

Using the $\mathrm{SU}(n)$ coherent states, we can write a diagonal resolution for the identity in $\mathds{B}_{N}^{n}$:

\begin{equation}
  \intl_{w\,\in\,\mathds{C}^{n-1}} d\mu(w\cg,w)\;|w\ket\bra w|=\mathds{1};\qquad
  d\mu(w\cg,w)=\frac{\sigma(n)\,\dim(\mathbb{B}_{N}^{n})}{(1+w\cg w)^{n}}\prodl_{j=1}^{n-1}d^{2}w_{j}.
  \label{eq2p3}
\end{equation}

Here again we use the notation $d^{2}w_{j}=dx_{j}dy_{j}=\frac{dw_{j}dw\cg_{j}}{2i}$, where $x_{j} = \Re\left(w_{j}\right)$ and
$y_{j} = \Im\left(w_{j}\right)$. Notice that we can deform the integration domain by making $w_{j}$ and $w\cg_{j}$ independent
variables, which amounts to the complexification of $x_{j}$ and $y_{j}$. We use this procedure in the following subsections to
replace the complex plane by a surface of integration in $\mathds{C}^{2}$ for each value of $j$. Also note that the normalization
constant in \eqref{eq2p3} can be divided into the factors $\sigma(n)=\frac{(n-1)!}{\pi^{n-1}}$, which is independent of the total
particle number, and $\dim(\mathbb{B}_{N}^{n}) = \frac{(N+n-1)!}{N!(n-1)!}$, the dimension of accessible Hilbert space.

The choice of parameterization defined in \eqref{eq2p1} for the $\mathrm{SU}(n)$ coherent states is quite appropriate, since the
integration domain in \eqref{eq2p3} is the same of equation \eqref{eq1p2}. Moreover, except for the normalization factor, the
coherent states are entire functions of the variables $w$.

\subsection{$\mathrm{SU}(n)$ coherent state propagator}
\label{ssc:suncohstatpropag}

Similarly to equation \eqref{eq1p3}, we define the propagator in the $\mathrm{SU}(n)$ coherent state representation as the
transition probability between the initial coherent state $|w_{i}\ket$ and the final coherent state $|w_{f}\ket$ after a time
interval $\tau$:

\begin{equation}
  K(w\cg_{f},w_{i};\tau)=\bra w_{f}|e^{-iH\tau}|w_{i}\ket.
  \label{eq2p4}
\end{equation}

For simplicity, we have assumed again a time-independent Hamiltonian $H$. To write down a path integral representation for
$K(w\cg_{f}, w_{i}; \tau)$, we factorize the evolution operator into $M$ identical propagation subintervals of length
$\vep=\frac{\tau}{M}$. Then, we introduce an identity of the form \eqref{eq2p3} between each pair of factors in the discretized
evolution operator:

\begin{equation}
  K(w\cg_{f},w_{i};\tau)=\int\left[\prodl_{k=1}^{M-1}d\mu(\b{w}^{k},w^{k})\right]
  \prodl_{j=1}^{M}\bra {\b{w}^{j}}\cg|e^{-iH\vep}|w^{j-1}\ket.
  \label{eq2p5}
\end{equation}

In the previous equation we define the integration variables $w^{k} = w(t_{k})$ and $\b{w}^{k} = \b{w}(t_{k})$, with $t_{k} =
k\vep$ for $k=1,2,\ldots,(M-1)$. We also extend this notation to the propagator boundary conditions by setting $w^{0}=w(0)=w_{i}$
and $\b{w}^{M} = \b{w}(\tau) = w_{f}^{*}$. Note that we employ the notation $w\cg(t) \rightarrow \b{w}(t)$ since the beginning of
this section, already evidencing the future duplication of the classical phase space. However, $w^{k}$ and $\b{w}^{k}$ need to be
considered as independent integration variables only after the semiclassical approximation, when the integration domain is
deformed to a surface in the doubled phase space which contains the $k$-th point of the discretized classical trajectory.
Nonetheless, to maintain the consistency of notation in this procedure, notice that the identities $\b{w}^{0} = {w^{0}}\cg =
w\cg_{i}$ and $w^{M} = {\b{w}^{M}}\cg=w_{f}$ must always be preserved ($\b{w}^{0}$ and $\b{w}^{M}$ can not be made independent of
$w^{0}$ and $w^{M}$), since these quantities do not stand for integration variables, which can be designated to follow the
classical trajectory in the doubled phase space.

Considering that at the some point of the calculations we will take the limit $M\rightarrow\infty$, we can expand each factor in
the integrand of \eqref{eq2p5} to first order in $\vep$:

\begin{equation}
  \begin{aligned}
  \bra {\b{w}^{j}}\cg|e^{-iH\vep}|w^{j-1}\ket&\approx\bra {\b{w}^{j}}\cg|\mathds{1}-iH\vep|w^{j-1}\ket\\
  &\approx\bra {\b{w}^{j}}\cg|w^{j-1}\ket e^{-i\vep\mathcal{H}_{j,j-1}};
  \end{aligned}
  \label{eq2p1ad}
\end{equation}

\noindent where $\mcl{H}_{j,j-1}=\frac{\bra{\b{w}^{j}}\cg|H|w^{j-1}\ket}{\bra{\b{w}^{j}}\cg|w^{j-1}\ket}$, for $j=1,2,\ldots,M$.
Then, substituting the expression \eqref{eq2p1ad} in \eqref{eq2p5}, we can readily obtain:

\begin{equation}
  K(w\cg_{f},w_{i};\tau)=\int\left[\prodl_{k=1}^{M-1}d\mu(\b{w}^{k},w^{k})\right]
  \exp\left[\suml_{j=1}^{M}\Ln\bra{\b{w}^{j}}\cg|w^{j-1}\ket-i\vep\suml_{j=1}^{M}\mcl{H}_{j,j-1}\right].
  \label{eq2p6}
\end{equation}

Employing the equation \eqref{eq2p2}, the argument of the exponential in \eqref{eq2p6} takes the following form:

\begin{equation}
  \begin{aligned}
  i\til{S}_{d}&=
  \suml_{j=1}^{M}\left\{\frac{N}{2}\Ln\left[\frac{(1+\b{w}^{j}w^{j-1})^{2}}
  {(1+\b{w}^{j}w^{j})(1+\b{w}^{j-1}w^{j-1})}\right]-i\vep\mcl{H}_{j,j-1}\right\}\\
  &= iS_{d}-\frac{N}{2}\Ln\left[(1+|w\cg_{f}|^{2})(1+|w_{i}|^{2})\right].
  \end{aligned}
  \label{eq2p7}
\end{equation}

In order to achieve the correct continuous limit of $\tilde{S}_{d}$, it is imperative to note that generally $w(\tau)\neq
w^{M}=w_{f}$ and $\b{w}(0)\neq \b{w}^{0}=w\cg_{i}$, thus ensuring the continuity of the trajectories in the independent variables
$w(t)$ and $\b{w}(t)$. Although $\tilde{S}_{d}$ explicitly depends on $w^{M}$ and $\b{w}^{0}$, as we see in the first line of
\eqref{eq2p7}, it is possible to analytically isolate all dependence on these quantities in a single term, as shown in the last
line of \eqref{eq2p7}. The term $S_{d}$ represents the discretized form of the action functional\cite{Note4}, whose extremization
correctly gives the classical equations of motion in the continuous-time limit.

Also in this limit, we can assume that consecutive values of $w(t_{k})$ and $\b{w}(t_{k})$ are very close, forming a continuous
trajectory. Therefore, according to \eqref{eq2p2}, we can write the following identity, which is valid up to first order in
$|w(t_{k})-w(t_{k})|$ and $|\b{w}(t_{k})-\b{w}(t_{k})|$:

\begin{equation}
  \Ln\bra \b{w}\cg(t_{k})|w(t_{k-1})\ket\approx
  \frac{N}{2}\frac{[\b{w}(t_{k})-\b{w}(t_{k-1})]w(t_{k-1})-\b{w}(t_{k})[w(t_{k})-w(t_{k-1})]}{1+\b{w}(t_{k})w(t_{k-1})}.
  \label{eq2p8}
\end{equation}

Adding and subtracting the appropriate terms dependent on $w(\tau)$ and $\b{w}(0)$, we can easily take the continuous limit of the
action functional $S_{d}$ with aid of expression \eqref{eq2p8}:

\begin{equation}
  iS(w\cg_{f},w_{i};\tau)=\intl_{0}^{\tau}dt\left\{\frac{N}{2}
  \frac{\dot{\b{w}}w-\b{w}\dot{w}}{1+\b{w}w}-i\mcl{H}(\b{w},w)\right\}
  +\frac{N}{2}\Ln\left\{[1+w\cg_{f}w(\tau)][1+\b{w}(0)w_{i}]\right\}.
  \label{eq2p9}
\end{equation}

In the previous equation we also defined the effective Hamiltonian $\mcl{H}(\b{w},w)=\frac{\bra \b{w}\cg|H|w\ket}{\bra
\b{w}\cg|w\ket}$, which represents the continuous limit of $\mcl{H}_{j,j-1}$. The extremization of $S$, for fixed boundary
conditions and propagation period, provides the classical equations of motion. So, by making $\delta S=0$, we obtain\cite{Note5}:

\begin{equation}
  \begin{aligned}
  \dot{w}&=-\frac{i}{N}(1+\b{w}w)\left(\mathds{1}+w\times\b{w}\right)\frac{\del\mcl{H}}{\del\b{w}}
  =-i\Xi\frac{\del\mcl{H}}{\del\b{w}};\\
  \dot{\b{w}}&=\frac{i}{N}(1+\b{w}w)\left(\mathds{1}+\b{w}\times w\right)\frac{\del\mcl{H}}{\del w}
  = i\b{\Xi}\frac{\del\mcl{H}}{\del w}.
  \end{aligned}
  \label{eq2p10}
\end{equation}

Here we should remember again that the classical trajectory is contained in a doubled phase space, since $w(t)$ and $\b{w}(t)$
must be treated as independent variables, preventing the overdetermination of the equations of motion caused by the boundary
conditions $w(0)=w_{i}$ and $\b{w}(\tau)=w\cg_{f}$.

Analogously to \eqref{eq1p9}, we can write equations of motion for deviations from the classical trajectory. Thus, we redefine
the matrices $R_{jk}$ for our new dynamical variables:

\begin{equation}
  \left(\begin{array}{c}
  \delta\dot{w} \\ \delta\dot{\b{w}}
  \end{array}\right)=
  \left(\begin{array}{c c}
  -i\frac{\del}{\del w}\left[\Xi\frac{\del\mcl{H}}{\del\b{w}}\right] &
  -i\frac{\del}{\del\b{w}}\left[\Xi\frac{\del\mcl{H}}{\del\b{w}}\right]\\
  i\frac{\del}{\del w}\left[\,\b{\Xi}\frac{\del\mcl{H}}{\del w}\right] &
  i\frac{\del}{\del\b{w}}\left[\,\b{\Xi}\frac{\del\mcl{H}}{\del w}\right]
  \end{array}\right)
  \left(\begin{array}{c}
  \delta w \\ \delta\b{w}
  \end{array}\right)
  =\left(\begin{array}{c c}
  R_{11} & R_{12} \\ R_{21} & R_{22}
  \end{array}\right)
  \left(\begin{array}{c}
  \delta w \\ \delta\b{w}
  \end{array}\right).
  \label{eq2p11}
\end{equation}

Notice that the identity $R_{22}^{T}=-R_{11}$ is no longer valid.

\subsection{Semiclassical approximation}
\label{ssc:semiapprox}

Substituting the definition \eqref{eq2p7} in equation \eqref{eq2p6} and then taking the continuous limit, we obtain a new
expression for the $\mathrm{SU}(n)$ propagator:

\begin{equation}
  K(w\cg_{f},w_{i};\tau)=\int D\mu(\b{w},w)
  \exp\left\{iS(w\cg_{f},w_{i};\tau)-\frac{N}{2}\Ln\left[(1+|\b{w}_{f}|^{2})(1+|w_{i}|^{2})\right]\right\}.
  \label{eq2p12}
\end{equation}

The quantity $D\mu(\b{w},w)$ symbolizes the following infinite product of measures in the doubled phase space:

\begin{equation}
  D\mu(\b{w},w)=\lim_{M\rightarrow\infty}\prodl_{k=1}^{M-1}\left[
  \frac{\sigma(n)\dim(\mathbb{B}^{n}_{N})}{(1+\b{w}^{k}w^{k})^{n}}\prodl_{j=1}^{n-1}d^{2}w^{k}_{j}\right].
  \label{eq2p13}
\end{equation}

Now, we proceed to the semiclassical approximation by expanding the action up to second order around the classical trajectory,
which we denote by $w_{c}(t)$ and $\b{w}_{c}(t)$:

\begin{equation}
  S\approx S_{c}+\frac{1}{2}\delta^{2}S_{c};\qquad\delta S_{c}=0.
  \label{eq2p14}
\end{equation}

Substituting the above approximation in the propagator \eqref{eq2p12}, we obtain the semiclassical propagator in the
$\mathrm{SU}(n)$ coherent state representation:

\begin{equation}
  K_{sc}(w\cg_{f},w_{i};\tau)=
  \exp\left\{iS_{c}-\frac{N}{2}\Ln\left[(1+|\b{w}_{f}|^{2})(1+|w_{i}|^{2})\right]\right\}
  \int D\mu(\b{w},w)\;e^{\frac{i}{2}\delta^{2}{S_{c}}}.
  \label{eq2p15}
\end{equation}

Therefore, it remains only to calculate the reduced propagator, defined by:

\begin{equation}
  K_{red}(w\cg_{f},w_{i};\tau)=\int D\mu(\b{\eta},\eta)\;e^{\frac{i}{2}\delta^{2}{S}_{c}};
  \label{eq2p16}
\end{equation}

\noindent where we have introduced new integration variables $\eta(t)=w(t)-w_{c}(t)$ and $\b{\eta}(t)=\b{w}(t)-\b{w}_{c}(t)$,
which represent deviations from the classical trajectory. Note that these new variables are subject to the boundary conditions
$\eta(0)=\b{\eta}(\tau)=0$. Expanding the action \eqref{eq2p9}, we get the following quadratic terms in $\eta$ and $\b{\eta}$
after some integrations by parts:

\begin{equation}
  i\delta^{2}S_{c}=\intl_{0}^{\tau}dt\;\left\{\eta\b{\Theta}\dot{\b{\eta}}-\b{\eta}\Theta\dot{\eta}
  +\eta A\eta+2\eta B\b{\eta}+\b{\eta}C\b{\eta}\right\}.
  \label{eq2p17}
\end{equation}

The matrices $\Theta$ and $\b{\Theta}$ are the inverses of $\Xi$ and $\b{\Xi}$, respectively. The explicit form of these matrices
is given by:

\begin{equation}
  \Theta=N\frac{(1+\b{w}w)\mathds{1}-w\times\b{w}}{(1+\b{w}w)^{2}}=\b{\Theta}^{T}.
  \label{eq2p18}
\end{equation}

The other quantities introduced in equation \eqref{eq2p17} are defined as follows:

\begin{equation}
  \begin{aligned}
  A&=\frac{2(\dot{\b{w}}w)\b{w}\times\b{w}-(1+\b{w}w)(\b{w}\times\dot{\b{w}}+\dot{\b{w}}\times\b{w})}
  {(1+\b{w}w)^{3}}-\frac{i}{N}\frac{\del^{2}\mcl{H}}{\del w^{2}};\\
  B&=\frac{1}{2}\frac{(\dot{\b{w}}w-\b{w}\dot{w})[2\b{w}\times w-(1+\b{w}w)\mathds{1}]
  +(1+\b{w}w)(\b{w}\times\dot{w}-\dot{\b{w}}\times w)}{(1+\b{w}w)^{3}}
  -\frac{i}{N}\frac{\del^{2}\mcl{H}}{\del w\del\b{w}};\\
  C&=\frac{(1+\b{w}w)(w\times\dot{w}+\dot{w}\times w)-2(\b{w}\dot{w})w\times w}
  {(1+\b{w}w)^{3}}-\frac{i}{N}\frac{\del^{2}\mcl{H}}{\del\b{w}^{2}}.
  \end{aligned}
  \label{eq2p19}
\end{equation}

Now we make a further change of integration variables, in order to compare the path integral \eqref{eq2p16} with the canonical
result \eqref{eq1p13}. To this end, we define the following matrices:

\begin{equation}
  Q=\sqrt{N}\frac{w\times\b{w}+\sqrt{1+\b{w}w}\left[(\b{w}w)\mathds{1}
  -w\times\b{w}\right]}{\b{w}w(1+\b{w}w)}=\b{Q}^{T};
  \label{eq2p20}
\end{equation}

\noindent such that $Q^{2}=\Theta$ and $\b{Q}^{2}=\b{\Theta}$. Then, applying the linear transformations $\nu=Q\eta$ and
$\b{\nu}=\b{Q}\b{\eta}$ in \eqref{eq2p17}, we get:

\begin{equation}
  i\delta^{2}S_{c}=\intl_{0}^{\tau}dt\;\left\{(\nu\dot{\b{\nu}}-\b{\nu}\dot{\nu})
  +\nu\til{A}\nu+2\nu\til{B}\b{\nu}+\b{\nu}\til{C}\b{\nu}\right\}.
  \label{eq2p21}
\end{equation}

Note that, in this new set of variables, $\delta^{2}S_{c}$ has the same form previously found in \eqref{eq1p11}. The
matrices above are related to variables of interest $w$ and $\b{w}$ as follows:

\begin{equation}
  \begin{aligned}
  \til{A}&=\b{Q}^{-1}AQ^{-1};\\
  \til{B}&=\b{Q}^{-1}B\b{Q}^{-1}-\frac{1}{2}\left(\dot{\b{Q}}\,\b{Q}^{-1}-\b{Q}^{-1}\dot{\b{Q}}\right);\\
  \til{C}&=Q^{-1}C\b{Q}^{-1}.
  \end{aligned}
  \label{eq2p22}
\end{equation}

In the previous definitions we used the inverses of the matrices $Q$ and $\b{Q}$, which can be readily found:

\begin{equation}
  Q^{-1}=\frac{1+\b{w}w}{\sqrt{N}\b{w}w}
  \left\{w\times\b{w}+\frac{\left[(\b{w}w)\mathds{1}-w\times\b{w}\right]}{\sqrt{1+\b{w}w}}\right\}
  =\left(\b{Q}^{-1}\right)^{T}.
  \label{eq2p23}
\end{equation}

Under the same transformation applied to \eqref{eq2p21}, the path measure \eqref{eq2p13} takes the following form:

\begin{equation}
  D\mu(\b{\nu},\nu)=\lim_{M\rightarrow\infty}\prodl_{k=1}^{M}\left[\frac{(N+n-1)!}{N!N^{n-1}}
  \prodl_{j=1}^{n-1}\frac{d^{2}\nu^{k}_{j}}{\pi}\right].
  \label{eq2p24}
\end{equation}

Therefore, $D\mu(\b{\nu},\nu)$ takes the same form found in \eqref{eq1p7} only when $N\gg n$. That is, the transformation given
by \eqref{eq2p20} produces the desired result only for large number of particles (macroscopic limit). However, this is precisely
the expected situation for the effectiveness of the semiclassical approximation, considering the computational gain and the
quantitative agreement with respect to exact quantum results.

In order to go from equation \eqref{eq2p13} to \eqref{eq2p24} we employed the explicit expression for $\det Q$, which is
trivially related to $\det\Theta$. The determinant of $\Theta$ can be easily calculated from the identity:

\begin{equation}
  \det\left[(1+\b{w}w)\mathds{1}-w\times\b{w}\right]=(1+\b{w}w)^{n-2};\quad\mbox{for } n\geq2.
  \label{eq2p25}
\end{equation}

Substituting \eqref{eq2p21} and \eqref{eq2p24} in \eqref{eq2p16}, under the additional condition $N\gg n$, we can recast the
reduced propagator in a similar way to the equation \eqref{eq1p12}:

\begin{equation}
  K_{red}(w\cg_{f},w_{i};\tau)=\int D\mu(\b{\nu},\nu)\;\exp\left\{\frac{1}{2}
  \intl_{0}^{\tau}dt\;\left[(\nu\dot{\b{\nu}}-\b{\nu}\dot{\nu})
  +\nu\til{A}\nu+2\nu\til{B}\b{\nu}+\b{\nu}\til{C}\b{\nu}\right]\right\}.
  \label{eq2p26}
\end{equation}

The variables $\nu(t)$ and $\b{\nu}(t)$ can be interpreted as deviations from the classical trajectory described in a new set of
dynamical variables, such that $\nu(t)=v(t)-v_{c}(t)$ and $\b{\nu}(t)=\b{v}(t)-\b{v}_{c}(t)$. With this interpretation in mind,
we can directly compare the equations \eqref{eq1p13} and \eqref{eq2p26}, allowing us to solve the integral of the reduced
propagator in an immediate way:

\begin{equation}
  K_{red}(w\cg_{f},w_{i};\tau)=\exp\left\{-\frac{1}{2}\intl_{0}^{\tau}\Tr\left[\til{B}(t)\right]dt\right\}
  \sqrt{\det\left[\frac{\del\b{v}(0)}{\del\b{v}(\tau)}\right]}.
  \label{eq2p27}
\end{equation}

However, the above result is not yet the desired solution, because we still need to return to the original variables $w$ and $\b{w}$.
Comparing the transformation between the variables $\eta$ ($\b{\eta}$) and $\nu$ ($\b{\nu}$) with their respective definitions, it becomes
evident the relation $v(t)=Q(t)w(t)$ ($\b{v}(t)=\b{Q}(t)\b{w}(t)$). Therefore, we can write the following identity:

\begin{equation}
  \det\left[\frac{\del\b{v}(0)}{\del\b{v}(\tau)}\right]
  =\det\left[\b{Q}(0)\frac{\del\b{w}(0)}{\del\b{w}(\tau)}{\b{Q}}^{-1}(\tau)\right]
  =\left[\frac{1+\b{w}(\tau)w(\tau)}{1+\b{w}(0)w(0)}\right]^{\frac{n}{2}}
  \det\left[\frac{\del\b{w}(0)}{\del\b{w}(\tau)}\right].
  \label{eq2p28}
\end{equation}

In the last equation we used the expression for the determinant of $\b{Q}(t)$, which can be easily calculated from
\eqref{eq2p25}. Furthermore, considering the quantities defined in \eqref{eq2p11}, \eqref{eq2p18}, \eqref{eq2p19} and
\eqref{eq2p22}, we can also obtain the relation:

\begin{equation}
  \Tr\left[\til{B}(t)\right]=\Tr\left[\b{\Theta}^{-1}(t)B(t)\right]=\frac{1}{2}\Tr\left[R_{11}(t)-R_{22}(t)\right].
  \label{eq2p29}
\end{equation}

At last, substituting \eqref{eq2p28} and \eqref{eq2p29} into \eqref{eq2p27}, we find the final expression for the semiclassical
propagator of $N$ bosons in $n$ modes:

\begin{equation}
  K_{sc}(w\cg_{f},w_{i};\tau)=
  e^{i(S_{c}+I)-\frac{N}{2}\Ln\left[(1+|\b{w}_{f}|^{2})(1+|w_{i}|^{2})\right]}
  \sqrt{\left[\frac{1+\b{w}(\tau)w(\tau)}{1+\b{w}(0)w(0)}\right]^{\frac{n}{2}}
  \det\left[\frac{\del\b{w}(0)}{\del\b{w}(\tau)}\right]}.
  \label{eq2p30}
\end{equation}

In the previous equation, we introduced the correction term to the classical action in this quantization scheme:

\begin{equation}
  I=\frac{1}{4}\intl_{0}^{\tau}\Tr\left[
  \frac{\del}{\del\b{w}}\left(\b{\Xi}\frac{\del\mcl{H}}{\del w}\right)
  +\frac{\del}{\del w}\left(\Xi\frac{\del\mcl{H}}{\del\b{w}}\right)\right]dt.
  \label{eq2p31}
\end{equation}

In comparison with the results for the harmonic-oscillator coherent states, we can also relate the factor under the square root
symbol in \eqref{eq2p30} with a second derivative of the classical action:

\begin{equation}
  \left[\frac{1+\b{w}(\tau)w(\tau)}{1+\b{w}(0)w(0)}\right]^{\frac{n}{2}}
  \det\left[\frac{\del\b{w}(0)}{\del\b{w}(\tau)}\right]=
  [1+w\cg_{f}w(\tau)]^{\frac{n}{2}}[1+\b{w}(0)w_{i}]^{\frac{n}{2}}
  \det\left[\frac{i}{N}\frac{\del^{2}S_{c}}{\del w_{i}\del w\cg_{f}}\right].
  \label{eq2p32}
\end{equation}

\section{Conclusion}
\label{sc:conclu}

We presented in detail a derivation for the semiclassical propagator in the $\mathrm{SU}(n)$ coherent state
representation based on previous findings for the multidimensional harmonic-oscillator coherent states. The main advantage of our
method is that it can easily be extended to other groups of physical interest. The transformation given by \eqref{eq2p20}, which
allows a direct comparison between the $\mathrm{SU}(n)$ path integral and the results for the harmonic-oscillator coherent
states, enables the immediate calculation of the semiclassical propagator in a phase space with more complex topology. This
simple procedure can be readily generalized to a wide range of dynamical groups, thus avoiding an additional laborious derivation
of the semiclassical propagator path integral for each new choice of coherent state representation, as usually done in previous
publications\cite{Braun07b, Ribeiro06}.

The equation \eqref{eq2p30} has direct application to the dynamical analysis of $N$ bosonic particles in $n$ modes. It is
expected that the $\mathrm{SU}(n)$ semiclassical propagator provides accurate results, when compared to exact quantum
calculations, for sufficiently large values of $N$. However, we can safely suppose that the approximation prescribed by $K_{sc}$
exceeds significantly the accuracy of mean-field methods, which are usually employed in the treatment of bosonic dynamics, even
for a relatively small number of particles\cite{ViscondiPrep}. We also expect the semiclassical propagation to be computationally
more efficient, in relation to similar quantum calculations, for systems with many degrees of freedom (large values of $n$). In
this case, the computational cost of calculation for the involved classical quantities grows slower with the number of modes than
their quantum counterparts.

\begin{acknowledgments}
We acknowledge the financial support from CNPq and FAPESP, under grants No. 2008/09491-9 and 2009/11032-5.
\end{acknowledgments}

\end{document}